\documentstyle[aps,prb]{revtex}
\begin{document}
\input epsf.sty

\twocolumn[\hsize\textwidth\columnwidth\hsize\csname 
@twocolumnfalse\endcsname

\draft

\title{Critical dynamics of a spin-5/2 2D isotropic antiferromagnet}
\author{R.~J. Christianson, R.~L. Leheny\footnote{Present address: Johns 
Hopkins University, Department of Physics and Astronomy, Baltimore, MD 21218}, R.~J. Birgeneau
\footnote{Present address: Department of Physics, University of Toronto,
Toronto, Canada M5S 1A7}}
\address{Department of Physics,
Massachusetts Institute of Technology, Cambridge, Massachusetts 02139}
\author{R.~W. Erwin}
\address{Reactor Radiation Division, NIST, Gaithersburg, Maryland 20899} 

\date{\today}
\maketitle

\begin{abstract}
We report a neutron scattering study of the dynamic spin correlations in 
Rb$_2$MnF$_4$, a two-dimensional 
spin-5/2 antiferromagnet.  By tuning an external magnetic field
to the value for the spin-flop line,
we reduce the effective spin anisotropy to essentially zero, thereby 
obtaining a nearly ideal two-dimensional isotropic antiferromagnet.  
From the shape of the quasielastic
peak as a function of temperature, we demonstrate dynamic scaling for 
this system and find a value for the dynamical exponent $z$. We compare 
these results to theoretical predictions for the dynamic
behavior of the two-dimensional Heisenberg model, in which deviations 
from $z=1$ provide a measure of the corrections to scaling.

\end{abstract}

\phantom{.}
]

\narrowtext

Recently, interest in
the two-dimensional (2D) square-lattice Heisenberg antiferromagnet has 
intensified due in large part to the
discovery of high-temperature superconductivity in the doped lamellar 
cuprates and the subsequent realization of the near-ideal 2D Heisenberg nature of their
parent compounds.\cite{Manousakis91}  The nearest-neighbor
Heisenberg model is defined as:
\begin{equation}
H = J \sum_{<i,j>} {\bf S}_{i} \cdot {\bf S}_{j}
\end{equation}
where $J$ is the nearest neighbor coupling which is positive for an antiferromagnet.
Classically, ${\bf S}_{i}$ is a three component vector of magnitude $\sqrt{S(S+1)}$
representing the spin at site $i$, while
quantum mechanically ${\bf S}_{i}$ is the quantum spin operator.

As a result of a symbiotic interplay among theory, simulation, and 
experiment, great 
progress in understanding the instantaneous spin correlations of the
2D Heisenberg antiferromagnet has been made in recent years.
Chakravarty, Halperin and
Nelson (CHN)\cite{Chakravarty} developed an effective field theory
from which an exact low-temperature expression for the instantaneous correlation 
length, $\xi$, has been found \cite{Hasenfratz91}.  While this expression
agrees closely with experiments on spin-1/2 Heisenberg systems 
\cite{Greven95}, measurements on systems with
$S>1/2$ display strong deviations from the predicted behavior \cite{Greven95,Leheny99}.
Subsequent work \cite{Elstner95,Cuccoli96} has pointed towards a broad crossover
from classical behavior at high temperature to the renormalized 
classical regime where field theory is valid.  Recently, Hasenfratz \cite{Hasenfratz99}
incorporated cutoff effects into the field theory formalism to describe the behavior in 
this crossover region. 

The dynamics of the 2D Heisenberg antiferromagnet has likewise been the 
subject
of detailed theoretical work. Tyc, Halperin and Chakravarty (THC) \cite{Tyc89} combined
renormalization group analysis and dynamic scaling theory\cite{Hohenberg77} with 
simulations of the classical lattice rotor 
model to predict a form for the dynamic structure factor.  
Classical molecular dynamics\cite{Wysin90} and quantum Monte Carlo simulations 
\cite{Makivic92} have lent credence to their predictions; however, due to a lack 
of suitable systems, comparatively few 
experimental studies of dynamics in 2D Heisenberg antiferromagnets have been
performed.  Some of the
most ideal 2D Heisenberg systems (La$_2$CuO$_4$ and Sr$_2$CuO$_2$Cl$_2$) have a
very large 
intersite coupling $J$ ($J \approx 1500K$), making quantitative
results using conventional neutron scattering techniques difficult to
obtain.  Consequently, previous experiments have not  
resolved the quasielastic scattering from the long-wavelength spin-wave
excitations.\cite{Yamada89,Thurber97}.  
In this communication,  we present a neutron scattering 
study of Rb$_2$MnF$_4$, a quasi-two-dimensional spin-5/2 system with an effective
spin anisotropy 
that can be tuned to zero using an external magnetic field. 
Our results provide a detailed
characterization of the dynamic structure factor in the quasielastic region.
Previous studies\cite{Leheny99,Birgeneau70} indicate that this system behaves 
like a nearly
ideal 2D Heisenberg antiferromagnet.  Accordingly, we compare our findings 
with the current theoretical understanding of 2D Heisenberg critical dynamics.

Following a strategy introduced in our previous work \cite{Leheny99}, we 
exploit the 
presence of a bicritical point in the field-temperature phase diagram
of Rb$_2$MnF$_4$ to make possible a study of the dynamic spin correlations
of a near-ideal Heisenberg system over a large range of correlation lengths.
Rb$_2$MnF$_4$ has the tetragonal $K_2NiF_4$ crystal structure with an
in-plane lattice constant of $a=4.215$ \AA \  and an out-of-plane lattice constant
of 13.77 \AA.  The large ratio of the out-of-plane to the in-plane lattice 
constant combines with the frustration due to the body-centered
stacking to make it a nearly two-dimensional magnetic system, with an 
interplane coupling of less than $10^{-4}J$.  

At zero field, Rb$_2$MnF$_4$ is a weakly Ising antiferromagnet 
with $J=0.63$ meV\cite{Cowley77}.  This interaction energy $J$ is more than two 
orders of magnitude smaller than that of the lamellar copper oxide
Heisenberg systems, thus making the energy scale of the dynamics much more accessible
for neutron scattering studies.  
The principal spin anisotropy is a magnetic dipole interaction, with 
$g\mu_BH_A = 0.032$ meV\cite{Cowley77} along the 
$c$ axis (perpendicular to the magnetic plane).
Correspondingly, when a field of approximately 5.5 T (depending on temperature) is 
applied parallel to the $c$ axis, the spins flop into the plane.  
Above this spin-flop transition, the
system has XY symmetry.  Precisely along the spin-flop 
line, and on the extension of the line into the paramagnetic 
phase, the anisotropy is effectively zero, so that the system should be
in the 2D Heisenberg universality class. (See Fig.\  1.)

Experiments were conducted at the NIST Center for Neutron Research
using NIST's 7 Tesla superconducting magnet.  We aligned the $c$ axis within
$0.5^{\circ}$ of the magnetic field to minimize any induced in-plane
anisotropies.  We took field scans at several temperatures to confirm the phase 
diagram and 
found the line of zero anisotropy in Tesla (shown in Fig.\  1) to be approximately:
$H = \sqrt{28.09 + 0.23 T}$ where $T$ is the temperature in Kelvin. This is
in accordance with the form given by Cowley {\em et al.} \cite{Cowley93}.  

Studies of the quasielastic scattering were performed with the 
thermal neutron triple-axis spectrometer BT9 and the cold neutron  
spectrometer SPINS.  At BT9, we used
a fixed initial energy of either 13.7 or 14.8 meV with a pyrolytic graphite filter
 before the sample to remove higher harmonics in the incident
beam.  Collimations of 40'-27'-Sample-24'-60' were typical, giving an energy 
resolution of 0.8 meV full width at half maximum (FWHM).  For lower 
temperatures where higher resolution was needed, we used SPINS with 
a fixed final energy of 4 meV and collimations of
guide-20'-S-20'-open, which gave a resolution of 0.12 meV FWHM.

Figure 2 shows the scattered intensity as a function of
energy at the antiferromagnetic zone center for several temperatures.
The two-dimensional Heisenberg antiferromagnet, in accordance with the 
Hohenberg-Mermin-Wagner
theorem, has no transition to long-range order above zero temperature.  
At non-zero temperature it has correlated regions whose characteristic length
scale diverges exponentially with inverse temperature.  These correlated
regions have a finite lifetime which 
translates into a non-zero energy width of the quasielastic peak
produced in the dynamic structure factor.  As the temperature 
is lowered towards zero, the correlated regions become progressively more stable,
and the energy width of the peak decreases.  The measurements in Fig.\  2
display this critical slowing down.

According to dynamic
scaling theory, the functional form of the structure factor is independent of 
temperature.  The temperature dependence enters only through the reduced reciprocal
space position $k$ (2D reciprocal space distance from the magnetic zone center) and 
frequency $\omega$, which are scaled by $\xi$, the correlation
length, and $\omega_0$, the characteristic frequency, respectively \cite{Hohenberg77}:
$q \equiv k\xi$ and $\nu \equiv \omega/\omega_0$, so that $q$ and $\nu$ are both
dimensionless.
In addition, the characteristic frequency is predicted to scale with 
the correlation length to a power $-z$, with $z=d/2$ for a Heisenberg 
antiferromagnet, where $d$ is the spatial dimension.

In accordance with these predictions, we fit the energy scans through the 
quasielastic peak at (0 1 0) to the dynamic structure factor:
\begin{equation}
S(k,\omega) = \omega_{0}^{-1} S(q) \Phi(q,\nu).
\end{equation}
We took Lorentzian forms for $S(q)$ and $\Phi(q,\nu)$:
\begin{eqnarray}
S(q) = \frac{S_0}{1 + q^2}
\\
\Phi(q,\nu) = \frac{\gamma_q^{-1}}{1 + \frac{\nu^2}{\gamma_q^2}}  
\end{eqnarray}
with $\gamma_q = (1 + \mu q^2)^{1/2}$.  ($\mu$ is an 
arbitrary constant.)  Due to the finite resolution, $q$-dependent contributions 
were needed to reproduce accurately
the observed lineshape.  $\mu = 1.7 \pm 0.2$ gave the best fit at all
temperatures, in agreement with the values ($1.4$ and $2.0$) found by THC 
in their analyses of the
classical lattice rotor model.  For the fits, $\xi$ was fixed at the values
determined in our previous study\cite{Leheny99}, and the temperature dependence 
of $S_0$ was found
to agree well with the results of those measurements.  The correlation length for the 
temperature range
accessible for studying the dynamics varied from $1<\xi/a<60$.

As Fig.\  2 demonstrates, fits to this form convolved with the experimental
resolution are quite good at all temperatures. Figure 3
shows the results for the energy widths extracted from these fits.  Note that
data taken at SPINS and BT9 agree closely in the overlapping
region. This agreement gives us confidence that we have correctly accounted
for the very different experimental resolutions in the two measurements. \cite{Note}
When the temperature is scaled by $JS(S+1)$, the temperature dependence
of $\omega_0(T)$ agrees well with the results of classical 
molecular dynamics simulations carried out 
by Wysin and Bishop \cite{Wysin90}.  They predict a temperature scaling
factor of $JS^2$, but when scaled by this factor, our data show a much
stronger temperature dependence than that exhibited by the simulation 
data.  Normalizing temperature by the
classical spin stiffness $JS(S+1)$ has been shown \cite{Elstner95} to 
collapse the instantaneous
correlation length data for 2D quantum Heisenberg antiferromagnets with $S>1$
onto the classical results at high temperatures,
and here again succeeds in reconciling the spin-5/2 data with 
corresponding results for the classical
system.  Thus, as with $\xi$, the dynamic behavior follows
classical scaling at high temperature ($1<\xi/a<10$). 

Similar measurements of the quasielastic energy width have been performed by 
Fulton {\em et al.}
\cite{Fulton94} on KFeF$_4$, another 2D spin-5/2 antiferromagnet.  These results 
agree with our data 
at the highest temperatures, but deviate strongly at lower temperatures.  We believe 
that this discrepancy results from a crossover to Ising 
critical behavior in KFeF$_4$.  Studies of Rb$_2$MnF$_4$ in zero field \cite{Lee98} 
show that the Ising crossover occurs near $1.2 T_N$.  
KFeF$_4$ has nearly the same reduced Ising anisotropy as Rb$_2$MnF$_4$,
and hence would also be expected to enter a region of Ising critical
behavior below $1.2 T_N$.  All but the highest 
temperatures from the KFeF$_4$ study therefore lie below the Ising crossover
region.

As mentioned above, dynamic scaling theory predicts $\omega_0 \propto \xi^{-z}$
with $z=1$ for the 2D Heisenberg antiferromagnet \cite{Hohenberg77}.  
In addition, CHN predict corrections to scaling which go as $T^{1/2}$:
$\omega_{0} = c\xi^{-1}(\frac{T}{2\pi\rho_s})^{1/2}$
where $c$ is the spin-wave velocity, and $\rho_s$ is the spin stiffness. 
Figure 4a shows a 
plot of $\omega_0$ versus $1/\xi$;  the best fit to the simple form
$\omega_0 \propto \xi^{-z}$ gives $z=1.35 \pm 0.02$.  This value for $z$ is
intermediate between the values for the 2D ($z=1$) and 3D ($z=1.5$) Heisenberg
antiferromagnets.  However, as detailed below we believe only a 2D model is
relevant here.  Figure 4b, which 
shows the product $\omega_0\xi$ versus temperature, demonstrates the
corrections to scaling if $z=1$ is assumed.  Clearly, corrections are
stronger than the $T^{1/2}$ predicted by CHN.  
Simulations of the classical model, as mentioned above, agree with our data for 
$\omega_0$, yet they claim to see a different temperature
dependence for the product $\omega_0\xi$.  This is most likely
due their use of a form for $\xi(T)$ that has since been shown to be inaccurate in
this temperature range. Monte Carlo studies on a spin-1/2
system \cite{Makivic92} have also indicated that $z=1$, but with a temperature
independent prefactor in agreement with predictions by Arovas and Auerbach.
\cite{Arovas88}

The data in Figs.\  3 and 4, taken at face value, may suggest that $z>1$ for 
Rb$_2$MnF$_4$ along the bicritical
line, or that there is a crossover to some other critical behavior with
a non-zero phase transition temperature.  
Explanations involving a crossover to three-dimensional behavior seem unlikely
in light of previous studies\cite{Birgeneau70,Lee98} at zero field showing that 
Rb$_2$MnF$_4$ behaves as a nearly ideal two dimensional system to very large 
correlation lengths.  Likewise, 2D Ising or 2D XY behavior are precluded by 
the high temperature at which we observe deviations from $z=1$
behavior, as compared to the scales at which these crossovers should occur,
as well as by our previous results on the statics\cite{Leheny99}, which agree 
very well with theory and simulation for the 2D Heisenberg model.

However, the dynamic scaling near the bicritical point could still conceivably 
differ from that of the 
ideal 2D Heisenberg antiferromagnet.  While the universality 
class for static critical behavior is determined solely by the symmetry 
properties and the spatial dimension, the dynamics can also be affected by
conserved quantities and the Poisson-bracket relations they satisfy \cite{Hohenberg77}.
The bicritical region differs from a true isotropic system due to the non-zero, 
conserved uniform magnetization along the applied
field direction.  Noting this distinction, Dohm and Janssen \cite{Dohm} performed a 
renormalization
group study of bicritical dynamics in $4-\epsilon$ dimensions.
They found that dynamic scaling was obeyed, but that the exponent for
the 3D bicritical point was larger than that for the 3D Heisenberg
model. To explore the possibility that we might be seeing a similar effect in
our 2D system, we measured $\omega_0$ in zero
field at temperatures above the Ising-Heisenberg crossover.  These results, 
shown in Fig.\  4a,
overlap closely with the data taken at the same temperatures on 
the bicritical line.  This indicates that, for this temperature
range, the magnetic field itself is not measurably affecting the quasielastic width.  
Clearly, additional
theoretical work on 2D bicritical dynamics and corrections to dynamic scaling for
the 2D Heisenberg antiferromagnet would greatly elucidate the findings from
these measurements.

With these measurements of the dynamic spin correlations in Rb$_2$MnF$_4$ near the
bicritical point, we have provided the first experimental study of
the quasielastic behavior in a 2D isotropic
antiferromagnet.  These results are largely consistent with
the current theoretical understanding of the dynamics of the 2D
Heisenberg model, but also raise some questions.  The shape of the dynamic structure
factor in the quasielastic region obeys a form consistent with dynamic scaling, and
the temperature dependence of the characteristic frequency $\omega_0$ is consistent
with the anticipated form $\xi^{-z}$, though with $z$ larger than the 
predicted value $z=1$.  To establish whether 
the difference in $z$ originates in stronger corrections to scaling than 
predicted or 
indicates a distinction between ideal 2D Heisenberg dynamic scaling and the dynamic
behavior near a 2D bicritical point will require
further theoretical work as well as experimental studies of other
Heisenberg antiferromagnets.  

\acknowledgements{We thank S.-H. Lee 
for his help with the experiments on SPINS.  This work was supported by the 
NSF Low Temperature Physics Program award number DMR 0071256.}

\listoffigures

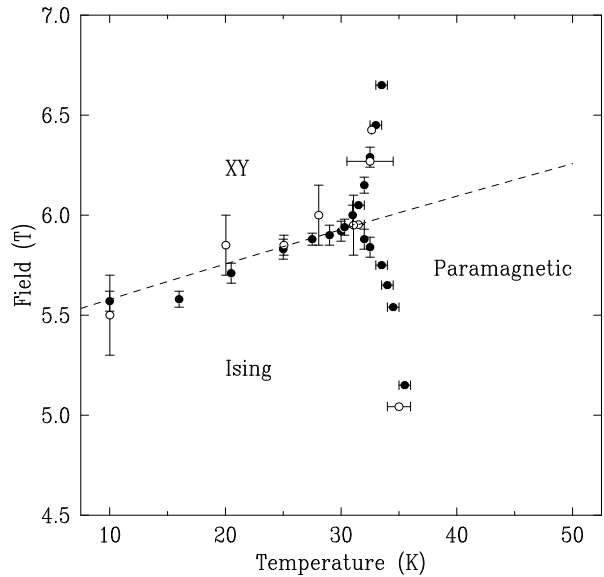
\begin{figure} 
\centerline{\epsfxsize=3.1in\epsfbox 
{Fig1.epsi}}
\caption[Phase diagram for Rb$_2$MnF$_4$ in an external magnetic field perpendicular
to the magnetic planes.  Open symbols are our measurements of the phase boundaries;
filled symbols are measurements from Cowley {\em et al.} \cite{Cowley93} shifted
by +0.15 T\@.  The dashed line indicates the line of zero effective anisotropy.]{}
\end{figure}

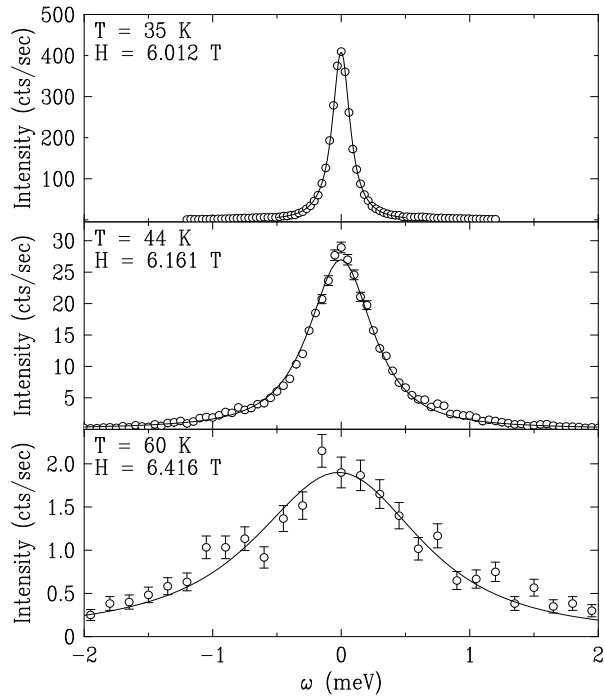
\begin{figure} 
\centerline{\epsfxsize=3.1in\epsfbox 
{Fig2.epsi}}
\caption[Energy scans through the quasielastic peak at the antiferromagnetic zone
center (0 1 0) at field and temperature values along the zero anisotropy line.
The solid lines show fits to Eqns. 2-4.  Scans shown were taken at BT9.]{}
\end{figure}

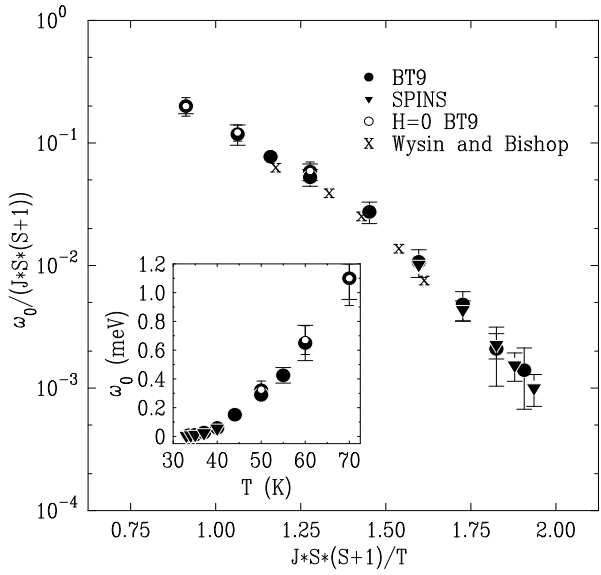
\begin{figure}  
\centerline{\epsfxsize=3.1in\epsfbox 
{Fig3.epsi}}
\caption[Measured energy width of the (0 1 0) quasielastic peak as a function
of temperature scaled by JS(S+1).  Classical simulation data by Wysin and Bishop has
been multiplied by an arbitrary constant as per their paper.  The inset shows our 
raw data.]{}
\end{figure}

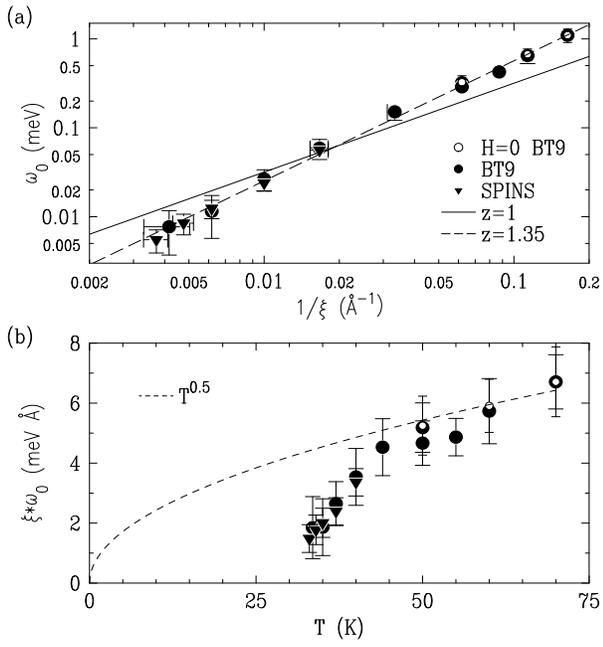
\begin{figure} 
\centerline{\epsfxsize=3.1in\epsfbox 
{Fig4.epsi}}
\caption[(a) $\omega_0$ versus $\xi$ showing best fits to z= 1 and 1.35 to the
entire range of data. (b) $\omega_0 \xi$ versus T to illustrate remnant
temperature dependence if $z$ is assumed to be 1.  The dashed line shows
the $T^{0.5}$ corrections to scaling predicted by CHN.]{}
\end{figure}


\begin{references}

\bibitem{Manousakis91}
For a review, see M.~A. Kastner, R.~J. Birgeneau, G. Shirane, Y. Endoh, Rev. Mod. Phys. 
{\bf 70}, 897 (1998).

\bibitem{Chakravarty}
S.Chakravarty, B.~I. Halperin, and D.~R. Nelson, 
Phys. Rev. B {\bf 39},  2344 (1989).

\bibitem{Hasenfratz91}
P. Hasenfratz and F. Niedermayer, Phys. Lett. B {\bf 268},  231  (1991).

\bibitem{Greven95}
M. Greven {\em et al.}, Zeitschrift fur Physik B {\bf 96}, 465 (1995). 

\bibitem{Leheny99}
R.~L. Leheny, R.~J. Christianson, R.~J. Birgeneau, R.~W. Erwin, Phys. Rev. Lett. 
{\bf 82}, 418 (1999).

\bibitem{Elstner95}
N. Elstner {\em et al.}, Phys. Rev. Lett. {\bf 75}, 938 (1995).

\bibitem{Cuccoli96}
A. Cuccoli, V. Tognetti, R. Vaia, P. Verrucchi, Phys. Rev. Lett. {\bf 77}, 3439 (1996).

\bibitem{Hasenfratz99}
P. Hasenfratz.  Euro. Phys. J. B {\bf 13}, 11 (1999).

\bibitem{Tyc89}
S. Tyc, B.~I. Halperin and S. Chakravarty, Phys. Rev. Lett. {\bf 62} 835
(1989).

\bibitem{Hohenberg77}
P.~C. Hohenberg and B.~I. Halperin,  Rev. Mod. Phys.  {\bf49}, 435 (1977).

\bibitem{Wysin90}
G.~M. Wysin and A.~R. Bishop,  Phys. Rev. B {\bf 42}, 810 (1990).

\bibitem{Makivic92}
M. Makivic and M. Jarrell,  Phys. Rev. Lett. {\bf 68}, 1770 (1992).

\bibitem{Hayden91}
S.~M. Hayden {\em et al.}, Phys. Rev. Lett. {bf 67} 3622 (1991).

\bibitem{Yamada89}
K. Yamada {\em et al.}, Phys. Rev. B {\bf 40}, 4557 (1989).

\bibitem{Thurber97}
K.~R. Thurber {\em et al.},  Phys. Rev. Lett. {\bf 79}, 171 (1997).

\bibitem{Birgeneau70}
R.~J. Birgeneau, H.~J. Guggenheim, G. Shirane, Phys. Rev. B {\bf 1}, 2211 (1970).

\bibitem{Cowley77}
R.~A. Cowley, G. Shirane, R.~J. Birgeneau, H.~J. Guggenheim , Phys. Rev. B 
{\bf 15}, 4292 (1977)

\bibitem{Cowley93}
R.~A. Cowley {\em et al.}, Z. Phys. B {\bf 93}, 5 (1993).
Our results 
agree quantitatively  with this study of the phase diagram of Rb$_2$MnF$_4$ provided 
that we shift the bicritical point by about
0.15 T.  This shift likely arises from 
minor differences in sample quality 
and small differences in the calibrations of the magnets used for the 
experiments.


\bibitem{Note}
In addition, the sample was remounted in the magnet twice during the experiment.
The good reproducibility indicates that any small possible misalignments of the sample 
in the magnetic field do not adversely affect the data.

\bibitem{Fulton94}
S. Fulton, R.~A. Cowley, A. Desert, T. Mason, J. Phys.: Cond. Matt. {\bf 6}, 6679 (1994).

\bibitem{Lee98}
Y.S. Lee {\em et al.}, Euro. Phys. J. B, {\bf 5} 1, 15 (1998).

\bibitem{Arovas88}  D.~P. Arovas and A. Auerbach, Phys. Rev. B {\bf 38}, 316 (1988).

\bibitem{Dohm}
V. Dohm and H.~K. Janssen,  Phys. Rev. Lett. {\bf 39}, 946 (1977);

\end{references}
\end{document}